\documentclass[%
reprint, 
superscriptaddress,
amsmath,amssymb,
aps
]{revtex4-2}

\usepackage{wasysym}
\usepackage{hyperref}
\usepackage{lipsum}
\usepackage{csquotes}
\usepackage{graphicx}
\usepackage{dcolumn}
\usepackage{bm}

\usepackage{printlen} 
\usepackage{setspace}
\usepackage[capitalise]{cleveref}
\usepackage{physics}
\usepackage{xspace}
\usepackage{xcolor}

\newcommand{\suppmat}{Supp.~Mat.\xspace}
\newcommand{\supprefdumbbellgeometry}{Supp.~Fig.~S1\xspace}
\newcommand{\supprefmsdfrombirth}{Supp.~Fig.~S2\xspace}

\newcommand{\supprefallfordisk}{Supp.~Figs.~S2 to S4\xspace}

\newcommand{\vect}[1]{\mathbf{\boldsymbol{#1}}}

\newcommand{\intd}{\,\mathrm{d}}

\let\oldchi\chi
\let\chi\undefined
\DeclareRobustCommand{\chi}{{\mathpalette\irchi\relax}}
\newcommand{\irchi}[2]{\raisebox{\depth}{$#1\oldchi$}} 

\newcommand{\cs}{~}

\newcounter{myfigpanel}[figure]
\newcounter{myfigpanelonly}[figure]

\newcommand{\panelletter}[1]{\refstepcounter{myfigpanel}\label{#1}\refstepcounter{myfigpanelonly}\label{onlyletter:#1}\alph{myfigpanel}}
\newcommand{\panel}[1]{(\protect\panelletter{#1})}

\crefalias{myfigpanel}{figure}
\crefname{myfigpanelonly}{panel}{panels}

\let\origcaption\caption
\let\caption\undefined
\DeclareRobustCommand{\caption}[1]{\origcaption{\setcounter{myfigpanel}{0}\setcounter{myfigpanelonly}{0}#1}}

\newcommand{\MPIDS}{\affiliation{Max Planck Institute for Dynamics and Self-Organization, Göttingen, Germany}}
\newcommand{\IDCS}{\affiliation{Institute for the Dynamics of Complex Systems, Göttingen University, Göttingen, Germany}}
\newcommand{\RPCTP}{\affiliation{Rudolf Peierls Centre for Theoretical Physics, University of Oxford, Oxford OX1 3PU, United Kingdom}}
\newcommand{\equalcontribution}{\thanks{these authors contributed equally}}

\begin{document}

\title{Isovolumetric dividing active matter}%

\author{Samantha R. Lish}%
\equalcontribution
\MPIDS
\RPCTP
\author{Lukas Hupe}%
\equalcontribution
\MPIDS
\IDCS
\author{Ramin Golestanian}
\MPIDS
\RPCTP
\IDCS
\author{Philip Bittihn}
\email{philip.bittihn@ds.mpg.de}
\MPIDS
\IDCS

\begin{abstract}
We introduce and theoretically investigate a minimal particle-based model for a new class of active matter where particles exhibit directional, volume-conserving division in confinement while interacting sterically, mimicking cells in early embryogenesis.
We find that complex motion, synchronized within division cycles, displays strong collective effects and becomes self-similar in the long-time limit.
Introducing the method of normalized retraced trajectories, we show that the transgenerational motion caused by cell division can be mapped to a time-inhomogenous random walk with an exponentially decreasing length scale.
Analytical predictions for this stochastic process allow us to extract effective parameters, indicating unusual effects of crowding and absence of jamming.
Robustness of our findings against desynchronized divisions, cell size dispersity, and variations in confinement hints at universal behavior.
Our results establish an understanding of complex dynamics exhibited by isovolumentric division over long timescales, paving the way for new bioengineering strategies and perspectives on living matter.

\end{abstract}

\maketitle

Classical active matter models have uncovered universal emergent behaviors in nature---such as phase separation\cs\cite{Tailleur2008,Buttinoni2013}, pattern formation\cs\cite{Glock2019,Ziepke2022}, and flocking\cs\cite{Vicsek1995}---emerging from a limited set of physical principles\cs\cite{Bowick2022,Copenhagen2020,Saha2020,Aondoyima2023}. These theoretical paradigms, often inspired by living systems, have primarily focused on motile\cs\cite{gompper_2020_2020} and chemically active matter\cs\cite{Golestanian2019,Hyman2014}. Recently, proliferation has been explored as a source of non-equilibrium activity in growing and dividing systems \cs\cite{Gelimson2015,Ambrosi2019,Hallatschek2023}, either leading to volume expansion\cs\cite{Farrell2013,Wang2017,You2018,IsenseeStressAnisotropy2022} or balanced by removal\cs\cite{Ranft2010,Pollack2022}. However, the physics underlying volume-conserving division has not been well studied. Such division is \textit{reductional} and is biomedically relevant for metazoan embryogenesis in confinement\cs\cite{ofarrell2015growing,schindler2024collective,Kumar2015}, normal tissue development\cs\cite{mDevany2023}, and pathogenetic growth during cancer metastasis\cs\cite{Lai2020}. While signaling and gene regulation are known to contribute to robust multicellular coordination\cs\cite{valencia2024combinatorial,delile2017cell}, we focus on mechanical drivers and simple scaling laws, whose importance is being increasingly appreciated\cs\cite{Fickentscher2018,Xu2018,Salle2022,Monfared2023,Fickentscher2024,Shroff2024,piszker2024fusion}.

Division into progressively smaller units does not, \textit{a priori}, imply any motion. However, division \textit{in vivo} is driven by mitosis and cytokinesis, which exert directional forces on their environment. Consequently, we consider a model based on elongation forces and steric repulsive interactions. In confinement, these ingredients lead to local rearrangements in the absence of large-scale expansion flows, otherwise only found in proliferating systems with removal or non-proliferating systems\cs\cite{Ranft2010,Blauth2021,Bi2016,Oyama2019,Zhang2023}. By tracking each element of the fractionating matter across successive divisions (generations) for arbitrarily long times, we find universal features---despite finite particle lifetimes, changing number density, and decreasing length scales. Global parameters such as cell cycle synchronicity and crowding\cs\cite{MalmiKakkada2023,Olejarz2018,Salle2022,schindler2024collective} modulate these effective dynamics, which may be seen in the context of glassy dynamics or (un)jamming found in other proliferating systems\cs\cite{Ranft2010,MalmiKakkada2018,Tjhung2020,Oswald2017,Blauth2021}. Our statistical-mechanics characterization of isovolumetric dividing active matter thus offers a new framework for understanding self-organization during proliferation.

\begin{figure*}[ht]
\includegraphics{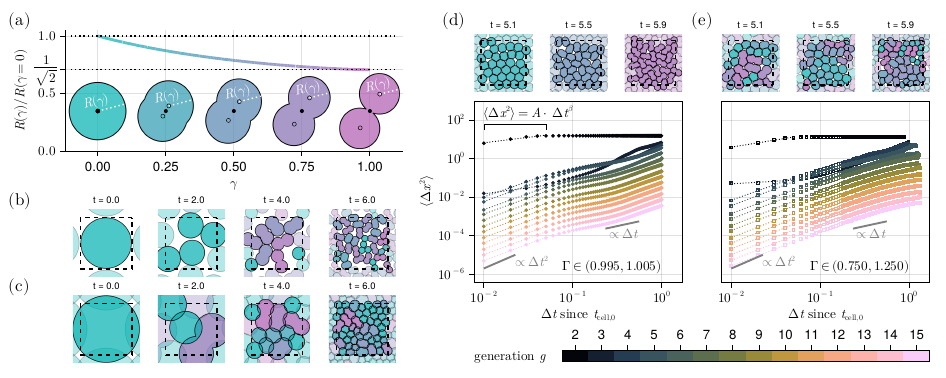}
\caption{Mean-squared displacements of cell centers during reductional division.
\panel{pan:shrinkingradius}~Reduction of the node radius to achieve volume conservation (see text), normalized here by the initial radius at birth $R_0=R(\gamma=0)$. Insets illustrate cell shape change resulting from simultaneous elongation and radius reduction from birth ($\gamma=0$) to division ($\gamma=1$).
\panel{pan:snapshotssparse}~Sample collective dynamics with steric interactions in a finite domain with periodic boundary conditions. Area fraction measuring crowdedness $\mathfrak{C}=V_\text{cell}/V_\text{domain}=0.74$.
\panel{pan:snapshotsdense}~As in \cref{onlyletter:pan:snapshotssparse}, but more crowded with $\mathfrak{C}=1.27$. Both examples: single-cell division rates sampled uniformly from the interval $(0.75, 1.25)$, new elongation axis chosen randomly for each daughter cell, periodic images of cells with center of mass outside the domain shown in lighter color.
\panel{pan:msdfrombirthnarrow}~Snapshots (top) and MSDs $\langle\Delta x^2\rangle$ (bottom) of cell centers, measured relative to the location of first appearance of each cell at time $t_{\text{cell},0}$. Narrow uniform distribution division rates $\Gamma$ across the interval $(0.995, 1.005)$.
\panel{pan:msdfrombirthwide}~Same as in \cref{onlyletter:pan:msdfrombirthwide}, but for wide uniform distribution of division rates $\Gamma\in (0.75, 1.25)$.
In \cref{onlyletter:pan:msdfrombirthnarrow,onlyletter:pan:msdfrombirthwide}, $\mathfrak{C}=1.01$ and different colors in MSD plot indicate different generations. Dashed lines towards the left indicate power-law fits of the form $\langle \Delta x^2\rangle=A\cdot \Delta t^\beta$ to the first 5 data points.
}
\label{fig:model_basicdynamics}
\label{fig:msdfrombirth}
\end{figure*}

\textit{Minimal model of isovolumetric division---}To investigate the collective dynamics of embryoid reductional division, we introduce volume conservation into a minimal model of dividing disk-shaped particles with smooth dynamics across cell divisions\cs\cite{diskcells2024arxiv}. The progression of each cell through its cell cycle is parameterized by an internal clock $\gamma\in [0,1)$ from birth to division with $\dot \gamma=\Gamma$. The division rate $\Gamma$ is drawn at birth, for each cell independently, from a uniform distribution. We investigate different widths of this distribution, always centered at 1, which defines our unit of time as a ``generation'' without loss of generality. Cells consist of two disks of radius $R$ whose centers are connected by an elastic spring with a rest length $2R\gamma$, i.e., fully overlapping at birth and just touching at division. $\Gamma$ therefore also acts as an elongation rate. The radius dynamically decreases with $\gamma$ according to %
$R(\gamma) = R_0\sqrt{\pi}/[2(\pi - \acos(\gamma) + \gamma\sqrt{1-\gamma^2})]^{1/2}$
to achieve exact volume conservation of the dumbbell shape which results from partially overlapping disks, as illustrated in \cref{pan:shrinkingradius} (see \suppmat and \supprefdumbbellgeometry for details). Daughter cells inherit the final radius of their mother---smaller by a factor of $1/\sqrt{2}$ in each generation---which ensures volume conservation of the entire population during successive embryonic cleavage divisions. The initial direction of elongation is chosen randomly at birth for each cell. The two disks of each cell interact with disks of other cells through Hertzian steric repulsion forces, which are scaled to ensure force continuity at division, when one disk is replaced by two completely overlapping ones. The equations of motion are then solved in the overdamped limit (see \cref{app:model} for details, initial conditions and parameters).

\textit{Complex cell motion---}Repeated reductional divisions create a system in which cellular matter continuously rearranges to accommodate the confinement, driven by cellular compartments which actively push away from each other to eventually form new cells (\cref{pan:shrinkingradius}) and steric repulsion between cells. How strongly cells interact during this process depends on the crowdedness $\mathfrak C$ of the environment, which is characterized here by the volume fraction $\mathfrak C = \sum_i V_\text{cell,i}/V_\text{domain}= V_\text{cell}/V_\text{domain}$, and constant in time due to volume-conserving division. For small $\mathfrak C$, the system is able to explore free space and relax mechanically (\cref{pan:snapshotssparse}), while forces between cells persist for large $\mathfrak C$, represented here by overlaps between their interaction boundaries (\cref{pan:snapshotsdense}).

To characterize the motion of cells, we start by tracking their center positions over time. Since cell centers do not exhibit continuous dynamics across divisions, we first restrict ourselves to displacements relative to the time of each cell's first appearance $\Delta t = t-t_{\text{cell},0}$. Mean-squared displacements (MSDs) $\langle\Delta x^2\rangle$ are then calculated by averaging over all cells within a generation $g$, since we expect systematic changes from generation to generation due to the reduction in particle size.

\Cref{pan:msdfrombirthnarrow,pan:msdfrombirthwide} show two cases (for $\mathfrak C$ close to 1) with narrow and wide division rate distributions, respectively. In the first case, divisions therefore happen rather synchronously (see snapshots at the top), whereas, in the second case, different cells are in different phases of their division cycle, leading to larger cell size dispersion. The resulting generationwise MSDs differ substantially in their shape for late generations: For narrow division rate distributions, MSDs increase rapidly both on very small sub-generational time scales and when approaching the cell cycle length, separated by a stagnation region. In contrast, these regimes are washed out by the desynchronization of growth cycles for wide division rate distributions, resulting in a steady decrease of MSD slope. However, both cases share a striking feature: While the MSDs of early generations are rather distinct, due to small-number effects and the shape of the confining domain (compare results for a circular physical container in \supprefmsdfrombirth), the curves for later generations only seem to differ in scale while converging to a universal shape.

\begin{figure}[ht]
\includegraphics{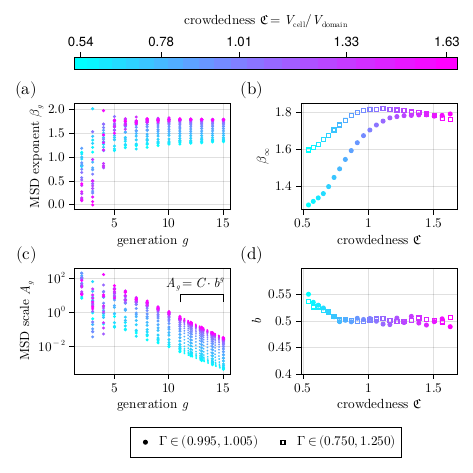}
\caption{Scaling behavior of cell-center MSDs, as indicated by the parameters $A$ and $\beta$ of the fitted power laws (cf. \cref{fig:msdfrombirth}).
\panel{pan:msdexponentgen}~Exponents $\beta_g$ as a function of generation $g$.
\panel{pan:msdexponentconf}~Limiting value of $\beta$ for large $g$ as a function of confinement $\mathfrak C$.
\panel{pan:msdscalegen}~Scale $A_g$ as a function of generation $g$.
\panel{pan:msdscalefactor}~Scaling factor $b=A_{g+1}/A_{g}$ between generations for large $g$ (as indicated by fit across the last 5 generations in \cref{pan:msdscalegen}) as a function of confinement $\mathfrak C$.
In all panels, different colors refer to different confinement strengths $\mathfrak C$ and symbols \scalebox{0.7}{$\newmoon$} and \scalebox{0.7}{$\square$} refer to narrow and wide division rate distributions, respectively (\cref{onlyletter:pan:msdexponentgen,onlyletter:pan:msdscalegen} show only data from the narrow distribution).
}
\label{fig:msdfrombirthscaling}
\end{figure}

To quantitatively assess this scaling behavior, we focus on the initial portion of the MSD curves, which corresponds to the initial mechanical relaxation behavior after division. By fitting a function $\langle \Delta x^2\rangle=A\cdot \Delta t^\beta$ to this part, we extract MSD exponents $\beta_g$ and scales $A_g$ for each generation $g$.
\Cref{pan:msdexponentgen} shows that, indeed, for large $g$, $\beta_g$ as a measure for the shape of the MSD curve, converges to a constant value $\beta_\infty=\lim_{g\rightarrow\infty}\beta_g$, which depends on crowdedness $\mathfrak C$. This dependence of the limiting exponent (\cref{pan:msdexponentconf}) again differs for different division rate distributions. For synchronously dividing cells, $\beta_\infty$ increases up to a limiting value at sufficiently high $\mathfrak C$. Increasing the density, and thus the mechanical coupling between cells, beyond this point does not change the scaling behavior of the initial relaxation phase anymore. In contrast, for more asynchronous cell cycles, non-monotonic behavior is observed and $\beta_\infty$ peaks around $\mathfrak C\approx 1$. This indicates an optimal mechanical coupling strength between cells causing particularly ballistic behavior, whereas coupling that is either too weak or too strong leads to slightly more diffusive motion.

It is worth noting that, in general, all motion performed by cell centers arises from interactions with other cells and is therefore of collective origin, as the elongation of an isolated dumbbell-shaped cell does not change its center position. Strong collective effects are also visible in the MSD scales $A_g$ (\cref{pan:msdscalegen}): With increasing volume fraction $\mathfrak C$, $A_g$ increases by roughly two orders of magnitude. Rather than causing the system to jam, higher density therefore leads to \emph{more} motion in the initial relaxation phase after division, as the system is fluidized by the introduction of new degrees of freedom due to reductional division. At fixed $\mathfrak C$, however, we again observe limiting behavior: For late generations $g$, $A_g$ shows almost perfect exponential scaling. The factor between generations (\cref{pan:msdscalefactor}) is universally $1/2$, except for very small $\mathfrak C$, i.e., sparse systems. This observation is remarkable given the strong collective effects observed before, but consistent with the reduction of cell radius by a factor $1/\sqrt{2}$ in each generation. Together with the converging MSD exponent (\cref{pan:msdexponentgen}), we have therefore established that, at late generations, when cells become significantly smaller than the domain size, cell motion becomes self-similar with an exponentially decreasing length scale that corresponds to cell size.

\begin{figure}[t]
\includegraphics[width=\columnwidth]{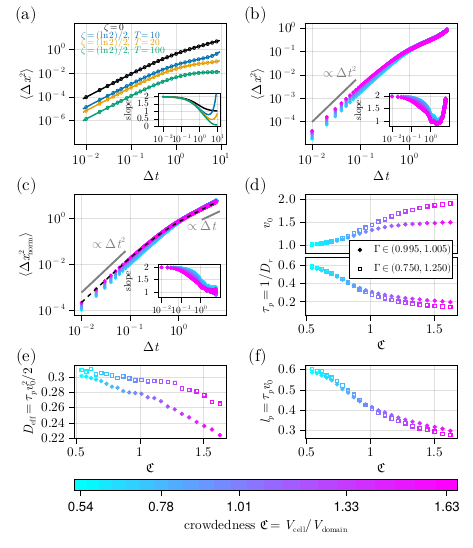}
\caption{Transgenerational compartment-based MSDs.
\panel{pan:abpmsd}~Time-averaged mean square displacement for ABP with exponential velocity suppression according to \cref{eq:abp}. Time-homogeneous case ($\zeta=0$, black) and three time-inhomogeneous cases with $\zeta=(\ln\,2)/2$ and different trajectory lengths $T$. Measured data from direct simulations of \cref{eq:abp} (symbols), theory from \cref{eq:abpmsd} (lines).
\panel{pan:nodemsdraw} Raw transgenerational MSDs of particle nodes for different values of $\mathfrak C = V_\text{cell}/V_\text{domain}$ (see colorbar).
\panel{pan:nodemsdnorm} MSDs of re-traced node trajectories normalized by the instantaneous node radius for different values of $\mathfrak C = V_\text{cell}/V_\text{domain}$ (see colorbar).
For \cref{onlyletter:pan:abpmsd,onlyletter:pan:nodemsdraw,onlyletter:pan:nodemsdnorm}, insets show the corresponding local log-log slopes indicating the apparent MSD exponent.
\panel{pan:fitparams} Parameters $v_0$ and $\tau_p = D_r^{-1}$ of least-squares fits of the theoretical mean-squared displacement without velocity suppression, $2v_0^2\tau_p[\tau_p(e^{-\Delta t/\tau_p} - 1) + \Delta t]$, to $\langle\Delta x_\text{norm}^2\rangle$ (exemplary fit shown as dashed line in \cref{onlyletter:pan:nodemsdnorm}).
\panel{pan:fitparams_Deff} Effective long-time diffusion constant $D_\text{eff}=\tau_p v_0^2/2$ and \panel{pan:fitparams_lp} persistence length $l_p=\tau v_0$ calculated from the effective ABP parameters in \cref{onlyletter:pan:fitparams}.
Symbols \scalebox{0.7}{$\newmoon$} and \scalebox{0.7}{$\square$} in all panels refer to narrow and wide division rate distributions, respectively (\cref{onlyletter:pan:nodemsdraw,onlyletter:pan:nodemsdnorm} show only data from the narrow distribution).
}
\label{fig:ensemblemsd}
\end{figure}

\textit{Time-inhomogenous persistent random walk analogy---}All analysis so far was limited to the dynamics within the limited lifetime of single cells, which is complex and highly parameter dependent (\cref{fig:msdfrombirth}). However, since the amount of material in the system is conserved, we can ask how individual parts behave on longer time scales. The idealized cells in our system are comprised of two compartments, which exhibit well-defined transgenerational dynamics. We therefore hypothesize that the motion of these individual elements can be understood as a persistent random walk with an exponentially decreasing length scale due to the size reduction of its constituent elements, which is consistent with the scaling behavior of collective effects characterized in \cref{fig:msdfrombirthscaling}. A simple model that produces such time-inhomogeneous random walks is a modified version of the well-known Active Brownian Particle (ABP) model\cs\cite{Howse2007,Debets2022}:
\begin{subequations}
\label{eq:abp}
\begin{align}
\label{eq:abppos}\dot{\vect r} &= v_0 e^{-\zeta t}\begin{pmatrix}
\cos(\theta) \\
\sin(\theta)
\end{pmatrix}\\
\label{eq:abporient}\dot\theta &= \chi.
\end{align}
\end{subequations}
The active velocity is $v_0$ at $t=0$, which sets the initial scale for the displacements, and then decreases exponentially with rate $\zeta$. The direction $\theta$ undergoes diffusive motion with $\langle\chi(t)\chi(t')\rangle=2D_r\delta(t'-t)$, where $D_r$ is the rotational diffusion constant. Note that, for simplicity, we have skipped the positional noise term. If it was included in \cref{eq:abppos}, its amplitude should also decrease in time in this picture. To match our particle-based model and the above scaling behavior, we assume that the typical length scale of motion decreases by a factor of $1/\sqrt 2$ in each generation, which we use as the unit of time, such that $\zeta=\frac{\ln 2}{2}$. The ensemble-averaged mean squared displacement with respect to the position at $t=0$ then evaluates to (see \cref{app:msd})
\begin{align}
\nonumber\langle&\Delta x^2\rangle = \langle ||\vect r(t)- \vect r(0)||^2\rangle\\
\label{eq:abpmsd}&= \frac{v_0^2}{D_r+\zeta} \left(\frac{2  e^{-t (\zeta +D_r)}+\frac{D_r}{\zeta} (1-e^{-2 \zeta  t}) -e^{-2 \zeta  t}-1}{D_r-\zeta}\right)
\end{align}
As expected, for small times, the MSD shows ballistic behavior $\propto v_0^2t^2$ and for $\zeta\rightarrow 0$, the expression converges to $2v_0^2\tau_p[\tau_p(e^{-t/\tau_p} - 1) + t]$, the well-known MSD of an athermal ABP with persistence time $\tau_p=D_r^{-1}$\cs\cite{Howse2007,Feng2017} (black line in \cref{pan:abpmsd}). For $t\rightarrow\infty$, the MSD approaches the finite value $v_0^2/[\zeta(D_r+\zeta)]$ if $\zeta\neq 0$ (see green line in \cref{pan:abpmsd}). Since instantaneous displacements resulting from \eqref{eq:abp} are non-stationary, additional corrections are necessary when time-averaging over trajectories of a finite length $T$ which is close to the maximum lag time $\Delta t$ (see \cref{app:finitetraj}). Direct numerical simulations of \eqref{eq:abp} confirm these calculations and show that it leads to an apparent increase in MSD slope towards large $\Delta t$ (see \cref{pan:abpmsd} and inset for $T=10$).

In our particle-based model of reductional division, we trace the entire set of all possible compartment-based trajectories from generation 9 onwards (see \cref{app:numanalysis} for details), when finite size effects have subsided according to \cref{pan:msdexponentgen,pan:msdscalegen}. The resulting time and ensemble-averaged MSDs are shown in \cref{pan:nodemsdraw} along with their scaling exponents. Indeed, we find qualitatively similar behavior as in \cref{pan:abpmsd}, including a clear drop of MSD exponents from initially 2 to below 1 and a rise towards the end. This prompts the question of whether the underlying compartment dynamics indeed exhibit an effective scaling analogous to the time-inhomogeneous random walk above, despite the combination of deterministic compartment separation and collective effects from cell-cell interactions. (The latter are visible in \cref{pan:nodemsdraw} as a dependency on the crowdedness $\mathfrak C$, which slightly increases MSDs at short times.) If such universal scaling behavior exists, we should be able to restore time-independence by applying a transformation
\begin{align}
\tilde{\mathbf{r}}_i(t) = \int_0^{t}f(t')\,\dot{\mathbf{r}}_i(t')\intd t'
\end{align}
to effectively eliminate the scaling and obtain \emph{normalized retraced} trajectories. Here, $\mathbf{r}_i(t)$ is a given measured trajectory and $f(t)$ is a transformation which scales the magnitude of incremental displacements in a time-dependent fashion. For the ABP analogy, $f(t)=e^{\zeta t}$ would trivially compensate the exponential velocity suppression in \cref{eq:abppos} and lead to the same MSDs as for an ordinary athermal ABP without velocity scaling ($\zeta=0$, black line in \cref{pan:abpmsd}). For the trajectories measured in our particle model of reductional division, a proxy for the instantaneous length scale of motion is the cell radius $R_i(t)$. Using $f(t)=1/R_i(t)$ to obtain normalized retraced trajectories and averaging over all data from generation 9 onwards leads to the MSDs shown in \cref{pan:nodemsdnorm}, indeed showing a classical transition from ballistic to diffusive behavior, which can be fitted with the theoretical MSD for $\zeta=0$, i.e., $2v_0^2\tau_p[\tau_p(e^{-t/\tau_p} - 1) + t]$ (see exemplary dashed line in \cref{pan:nodemsdnorm}).

Given the successful renormalization, we can now use these fits to extract effective parameters for the underlying microscopic process, which include the collective effects from steric interactions. As \cref{pan:fitparams} shows, compartments in sparse systems with low crowdedness $\mathfrak C$ have the largest persistence time $\tau_p$, which is consistent with their undisturbed motion and weak steric interactions. Similarly, an effective self-propulsion velocity $v_0$ of 1 cell radius per generation corresponds exactly to the bare velocity of a cell compartment. With increasing crowdedness, the self-propulsion velocity increases while the persistence time decreases, consistent with more frequent and intense perturbations by neighboring cells. The parameters for low $\mathfrak C$ are remarkably independent of the division rate distribution (i.e., synchronous vs. asynchronous divisions, compare \cref{fig:model_basicdynamics}). In contrast, for higher $\mathfrak C$, the trend towards faster but less persistent motion is clearly stronger for asynchronous cell divisions and thus larger cell size dispersion. Together, these parameters determine the long-time effective diffusion coefficient $D_\text{eff}=\tau_p v_0^2/2$ which generally decreases due to crowding (\cref{pan:fitparams_Deff}) as could be expected. Interestingly, for the broad division rate distribution, the increased heterogeneity in sizes and asynchronicity of divisions seems to partly make up for the crowding effects and the decrease is much less pronounced. In contrast, the persistence length $l_p=\tau_p v_0$ (\cref{pan:fitparams_lp}), which decreases towards higher $\mathfrak C$, is relatively insensitive to the width of the division rate distribution.

In this study, we have shown that rearrangements on increasingly finer scales, due to isovolumetric division and purely steric interactions, can be understood as an effective persistent random walk with a decreasing length scale. This mapping is possible despite strong collective effects, which depend on crowdedness and cell cycle synchronicity, and lead to corrections in the effective parameters.
Here, we focused on the long-time limit, when self-similarity emerges and the dynamics do not depend on the shape of the confinement (cf. results for a circular enclosure in \supprefallfordisk). However, boundaries clearly play a role only in the early phase, when the cell size is still comparable to domain shape features and the majority of cells still touch the boundary\cs\cite{SeirinLee2022,Shroff2024} (compare early generations \cref{fig:msdfrombirth} and \supprefmsdfrombirth).
In contrast to models aimed at reproducing exact division order and morphologies for low cell number during embryogenesis\cs\cite{Fickentscher2013,Yamamoto2017,Tian2020,Kuang2023,Koyama2023}, our results serve as a baseline identifying universal rules in isovolumetrically dividing systems. Future studies could make use of the identified scaling behavior to explore allometric properties of reductional division\cs\cite{Fickentscher2024,Salle2022,schindler2024collective} or to design reconfigurable self-assembled organisms\cs\cite{kriegman2020scalable}.

Interestingly, increasing crowdedness does not cause jamming in our system, but leads to stronger yet more random motion. This is due to the unconventional role of steric repulsion: While in principle being a passive (conservative) force, each division event untethers cellular compartments, creating new degrees of freedom and thereby opening previously inaccessible pathways for relaxation. Thus, the relaxation dynamics are stronger at higher densities, adding to the deterministic compartment separation inherent to each cell (cf. \cref{pan:fitparams}). Due to overall volume conservation, bypassing jamming is possible despite the absence of apoptosis, which has been observed to cause tissue fluidization together with cell division\cs\cite{Ranft2010,Oswald2017,Blauth2021}. It would be interesting to explore how different interaction potentials change the contribution of these relaxation dynamics to the effective parameters of the system.




It is worth noting that compartment separation in our model is designed to be smooth and spread out over the entire cell cycle---in contrast to other models with instantaneous replacement upon division\cs\cite{Tjhung2020}. In reality, mitosis and cytokinesis for biological cells only occupy certain phases of the cell cycle. However, on transgenerational time scales, we expect our results to be independent of the exact time course of compartment separation as long as it is sufficiently smooth.

Finally, retracing and normalizing trajectories in isovolumetric dividing matter revealed that seemingly anomalous behavior (MSDs in Figs.~\ref{pan:msdfrombirthnarrow},\ref{onlyletter:pan:msdfrombirthwide} \& \ref{pan:nodemsdraw}) can be caused by simpler underlying dynamics. Thus, our results demonstrate how traditional statistical physics measures can be generalized to extract information from self-similar systems with changing intrinsic scales despite emergent complexity. This could be important not only for the search of universality in theoretical models, but also inform appropriate interpretation of experimental data.\\

\begin{acknowledgements}
We thank Hari Shroff and his group at the Janelia Research Campus for inspirational discussions on the biophysics of embryogenesis, as well as Peter Sollich and Yoav Pollack for illuminating conversations.
Research reported in this publication was supported by the National Institute of Biomedical Imaging and Bioengineering (NIBIB), at the National Institutes of Health under the NIH Oxford-Cambridge Scholars Program (S.R.L).
\end{acknowledgements}

\section*{Software Availability}
An implementation of the reductional division model in the Julia programming language\cs\cite{bezanson_julia_2017} will be made available via the package \href{http://biome.inparts.org}{InPartSBiome.jl} at \url{http://biome.inparts.org}~\cite{InPartSBiome}, together with example simulation scripts.

\appendix
\section{Particle-based model and numerical analysis}
\label{app:model}
\label{app:numanalysis}
As described in the main text, our isovolumetric division model is based on a minimal model of smoothly dividing disk-shaped cells, which consist of two circular nodes that push apart driven by spring with changing rest length. The definitions of all degrees of freedom (cell position, orientation, rest length of the internal spring) and the equations of motion determining their time evolution are described in detail in Ref.~\citenum{diskcells2024arxiv}. The main feature of the model which is important in the context of this study is its mechanical consistency: All interaction laws are designed to achieve continuity of all forces on nodes even across divisions. The positions of the nodes also connect seamlessly to the node positions of the mother cell, yielding continuous trajectories across generations which enable the analysis in \cref{fig:ensemblemsd}.

The basic model from Ref.~\citenum{diskcells2024arxiv} is modified here in two ways: First, the orientation of each cell is chosen randomly at birth ($\gamma=0$), when it can be chosen freely without any mechanical penalty due to the circular shape of the cell at birth. This sets a well-defined time scale for orientational memory.
As a second difference, the node radius $R$ of a cell is not constant, but changes with the internal clock of the cell cycle, which we denote here by $\gamma\in [0,1)$ instead of $g$ as in Ref.~\citenum{diskcells2024arxiv}, since we use $g$ to denote the generation of a cell (accordingly, we use $\Gamma$ to refer to the division rate $\dot \gamma$ here). We calculate this instantaneous radius using
$R(\gamma) = R_0\sqrt{\pi}/[2(\pi - \acos(\gamma) + \gamma\sqrt{1-\gamma^2})]^{1/2}$ to ensure that the total two-dimensional volume of a dumbbell-shaped cell remains constant throughout its division cycle (see \cref{pan:shrinkingradius} and \supprefdumbbellgeometry), ending with $R(\gamma=1)=R_0/\sqrt{2}$ at division, which is the initial radius for the children. Note that we do use the implicit radius dependencies of both the force laws for steric (Hertzian) interactions and the mobilities\cs\cite{Luders2021} as detailed in Ref.~\citenum{diskcells2024arxiv} to achieve mechanical consistency. For the former, this also means taking into account potentially unequal radii of two interaction partners in the Hertzian force law, such that the magnitude of the interaction force between two nodes with radii $R_1$ and $R_2$ becomes
\begin{align}
|\vect{F}| = mY^*\sqrt{R^*}\left(R_1+R_2 - d\right)^{3/2}\text{ for }d \le R_1+R_2,
\end{align}
where $m$ is a softening factor introduced in Ref.~\citenum{diskcells2024arxiv} to achieve force continuity across divisions and the effective radius $R^* = (R_1^{-1}+R_2^{-1})^{-1}$. Since we use the same effective Young's modulus $Y$ for all cellular material, $Y^*$ still simplifies to $Y^*=(Y_1^{-1}+Y_2^{-1})^{-1}=Y/2$. Assuming overdamped dynamics, the (radius and $\gamma$ dependent) mobilities then translate total forces and torques to instantaneous velocities as detailed in\cs\cite{diskcells2024arxiv}.

Each simulation is set up with a circular single cell ($\gamma=0$) of radius $R_\text{init}$ in the domain center with a randomly chosen orientation that defines its initial direction of elongation. Due to exact volume conservation across individual divisions, the radius $R_\text{init}$ of this initial cell together with the domain volume $V_\text{domain}$ defines the crowdedness $\mathfrak C=\pi R_\text{init}^2/V_\text{domain}$ for the entire simulation. Note that we chose to keep $R_\text{init}$ constant between simulations and instead vary the domain volume to achieve different values of $\mathfrak C$. Since both the force laws as well as the mobilities of the particles are radius-dependent (see above), this makes sure that the dynamical parameters of individual cells remain unchanged when comparing different $\mathfrak C$.

We then allow this initial cell to divide exactly 15 times, leading to a final state with 32768 circular cells with radius $R_\text{init}/2^{15/2}$ just after reaching generation 16. Depending on the distribution from which the division rates $\Gamma_i$ for each cell $i$ are chosen, these divisions happen synchronously or asynchronously, as outlined in the main text. The entire system is simulated for a time span of 20 generations, when all cells have stopped dividing even for the widest division rate distribution. For MSDs calculated from cell birth within each generation, there are therefore 15 generations of available data (see, e.g., \cref{pan:msdfrombirthnarrow},\ref{onlyletter:pan:msdfrombirthwide}). For transgenerational compartment-based trajectories, we use data from generation 9 onwards as outlined in the main text, up to the point where the first cell has stopped dividing. This allows us to examine the ``long-time limit'' before the effect of finite-time simulations can be felt. Individual transgenerational trajectories are computed by tracing the continuous position of all final nodes through all divisions back to the beginning of generation 9. The increasing multiplicity of the data towards the beginning (due to forward-splitting/backward-merging trajectories) is not removed. Once these trajectories have been obtained, MSDs are calculated as usual via time and ensemble averaging. Statistics for all MSD calculations (both cell-based intragenerational and node-based transgenerational) is improved by pooling data from 10 independent realizations for each parameter set (computed with different random seeds).

Constant model parameters used in all simulations of this study are: initial radius $R_\text{init}=16$, viscosity for mobilities\cs\cite{Luders2021} $\eta=0.05$, effective Young's modulus for Hertzian repulsion $Y=200$. Snapshots of the system are taken every $0.01$ generations, setting the minimum time lag for MSD calculations.

\section{MSD for ABPs with exponential velocity suppression}
\label{app:msd}
Here, we calculate the mean squared displacement for an ABP with dynamics as in \cref{eq:abp}.
Because of isotropy, it is sufficient to calculate the $x$ component of the MSD:
\begin{align}
\nonumber |r_x(t)&-r_x(0)|^2 \\
\nonumber    &= \int_0^t\int_0^t v_x(u)v_x(u')\intd u \intd u'\\
\nonumber    &= 2\int_{u=0}^t\int_{u'=u}^t e^{-2\zeta u}v_x(0)v_x(u'-u)\intd u\intd u'\\
\label{eq:generalMSDfromVACF}    &= 2\int_{u=0}^t\int_{s=0}^{t-u} e^{-2\zeta u}v_x(0)v_x(s)\intd u\intd s,
\end{align}
where integration variable names were inserted at the lower integral bounds for clarity.
Taking the ensemble average on both sides, we need the velocity auto-correlation $\langle v_x(0) v_x(s)\rangle$, which is straight-forward to calculate using the probability distribution
\begin{align}
p(\theta,t)=\frac{1}{\sqrt{4\pi D_r t}} \exp\left(-\frac{(\theta -\theta_0)^2}{4 D_r t}\right),
\end{align}
which solves \cref{eq:abporient} for an initial angle $\theta_0$. This leads to $\langle v_x(0) v_x(s)\rangle = \exp\left[-t(D_r+\zeta)\right]v_0^2\cos^2(\theta_0)$ or $\exp\left[-t(D_r+\zeta)\right]v_0^2/2$ averaging over initial angles. Substituting this into the ensemble average of \cref{eq:generalMSDfromVACF} and using $\langle |\vect r(t)- \vect r(0)|^2\rangle = 2\langle|r_x(t)-r_x(0)|^2\rangle$ then leads to
\begin{align}
\nonumber\langle ||&\vect r(t)- \vect r(0)||^2\rangle\\
\label{eq:abpmsd_alt}&= \frac{v_0^2}{\zeta  (D_r + \zeta)} \left(1-\frac{e^{-2 \zeta  t} \left(D_r+\zeta -2 \zeta  e^{-t (D_r-\zeta)}\right)}{D_r-\zeta}\right)
\end{align}
Despite its lengthy form, it is easy to verify that $D_r=\zeta$ is a removable singularity. The limit $t\rightarrow\infty$ is easiest to obtain from the above expression, whereas the limit $\zeta\rightarrow 0$ of an ordinary athermal ABP is more obvious from the alternative form of \cref{eq:abpmsd}.

\section{Time-averaged MSD for finite data}
\label{app:finitetraj}
Due to the explicit time dependence of the effective self-propulsion velocity $v_0 e^{-\zeta t}$ in \cref{eq:abp}, the mean squared displacement of \cref{eq:abpmsd_alt} and \cref{eq:abpmsd} calculated in \cref{app:msd} is only valid when measured with respect to the position at time $t=0$. However, using a reference time point $t=t_0$ instead of $t=0$ merely rescales the initial self-propulsion velocity from $v_0$ to $e^{-\zeta t_0}v_0$. Making the same substitution in \cref{eq:abpmsd_alt} or \labelcref{eq:abpmsd} therefore easily translates the above result to an arbitrary reference point:
\begin{align}
\label{eq:msdrescaling_singletimepoint}\langle ||\vect r(t_0 + \Delta t)- \vect r(t_0)||^2\rangle = e^{-2\zeta t_0}\langle ||\vect r(\Delta t)- \vect r(0)||^2\rangle
\end{align}
In systems with time-translation symmetry, time averaging -- averaging the MSD measured from different reference time points within a time series -- is possible without distorting the results and is usually employed to improve statistics. In the present system, the same procedure has two effects: First, because the MSD is rescaled for larger $t_0$ according to \cref{eq:msdrescaling_singletimepoint}, averaging over longer time series will also lead to a decrease of the time-averaged MSD. Secondly, if the largest time lag $\Delta t_\text{max}$ is on the same order as the length $T$ of the time series, the possible values of $t_0$ depend on and are constrained by the value of $\Delta t$, since we require $t_0+\Delta t\leq T$. The total rescaling factor for a time series length $T$ and given $\Delta t$ can easily be calculated with the help of \cref{eq:msdrescaling_singletimepoint} as
\begin{align}
\nonumber F(T,\Delta t) &= \frac{1}{T-\Delta t}\int_0^{T-\Delta t}e^{-2\zeta t_0}\intd t_0\\
\label{eq:msdrescaling_timeaveragefactor}&= \frac{1-e^{-2\zeta (T-\Delta t)}}{2\zeta(T-\Delta t)}
\end{align}
The overall time-averaged and ensemble-averaged MSD from such time series of length $T$ will therefore be
\begin{align}
\langle \Delta x^2\rangle &= F(T,\Delta t)\cdot \langle ||\vect r(\Delta t)- \vect r(0)||^2\rangle
\end{align}
with $F$ from \cref{eq:msdrescaling_timeaveragefactor} and $\langle ||\vect r(\Delta t)- \vect r(0)||^2\rangle$ from \cref{eq:abpmsd} or \cref{eq:abpmsd_alt}.
It can easily be seen that, as expected, MSDs are simply rescaled by a common factor $\approx(1-e^{-2\zeta T})/(2\zeta T) < 1$ independent of $\Delta t$ as long as $\Delta t\ll T$. Slopes (i.e., MSD exponents) in this regime are therefore unaffected by the rescaling. For $\Delta t$ approaching $T$, however, we expect a ``false'' increase in MSDs and slope.


\end{document}